\documentclass[twocolumn,preprintnumbers,aps,prd,superscriptaddress,nofootinbib,floatfix,showpacs,showkeys]{revtex4-1}

\usepackage{braket}
\usepackage{float}
\usepackage{graphicx}  
\usepackage{csquotes}
\usepackage{multirow}
\usepackage{natbib}
\usepackage{pifont}
\usepackage{amssymb}
\usepackage{bm}        
\usepackage{amsfonts}  
\usepackage{amsmath}   
\usepackage{amssymb}   
\usepackage[titletoc]{appendix}
\usepackage{color}
\usepackage{hyperref}
\usepackage{cleveref}
\usepackage[english]{babel}
\usepackage{booktabs}
\usepackage{mathtools}
\usepackage{subfigure}
\usepackage{enumerate}

\definecolor{vividviolet}{rgb}{0.62, 0.0, 1.0}
\definecolor{amaranth}{rgb}{0.9, 0.17, 0.31}
\definecolor{palatinateblue}{rgb}{0.15, 0.23, 0.89}
\definecolor{brightpink}{rgb}{1.0, 0.0, 0.5}
\definecolor{cornflowerblue}{rgb}{0.39, 0.58, 0.93}
\definecolor{deepcarminepink}{rgb}{0.94, 0.19, 0.22}
\definecolor{radicalred}{rgb}{1.0, 0.21, 0.37}

\hypersetup{ linktoc=all,
    colorlinks, linkcolor={palatinateblue},
    citecolor={brightpink}, urlcolor={amaranth}
}

\def\be{\begin{equation}}
\def\ee{\end{equation}}

\begin{document}

\title{Multipartite entanglement features of primordial non-gaussianities}

\author{Alessio Belfiglio}
\email{alessio.belfiglio@unisi.it}
\affiliation{DSFTA, University of Siena, Via Roma 56, 53100 Siena, Italy.}

\author{Roberto Franzosi}
\email{roberto.franzosi@unisi.it}
\affiliation{DSFTA, University of Siena, Via Roma 56, 53100 Siena, Italy.}
\affiliation{Istituto Nazionale di Fisica Nucleare (INFN), Sezione di Perugia, Perugia, 06123, Italy.}

\author{Orlando Luongo}
\email{orlando.luongo@unicam.it}
\affiliation{University of Camerino, Via Madonna delle Carceri, Camerino, 62032, Italy.}
\affiliation{Department of Mathematics and Physics, SUNY Polytechnic Institute, Utica, NY 13502, USA.}
\affiliation{INAF - Osservatorio Astronomico di Brera, Milano, Italy.}
\affiliation{Al-Farabi Kazakh National University, Almaty, 050040, Kazakhstan.}

\begin{abstract} 
We discuss some entanglement features associated with cubic non-Gaussian perturbations in single-field inflationary scenarios. We adopt standard momentum-space techniques to show how multipartite entanglement arises for inflationary perturbation modes, focusing on the dynamics of the comoving curvature perturbation. In particular, we quantify entanglement generation via the recently proposed Entanglement Distance, which introduces a geometric interpretation of quantum correlations in terms of the Fubini-Study metric. In the continuum limit, we show that the Entanglement Distance arising from displacement transformations is proportional to the total number of excitations in the quantum state for cubic perturbations, thus providing an upper bound on the von Neumann entanglement entropy of any reduced state compatible with such excitations. Within the interaction picture, we further observe that the quantum correlations arising from cubic gravitational interactions are typically much larger than the standard squeezing contribution, in agreement with previous studies focusing on von Neumann entropy generation across the Hubble horizon. We further show how the inflationary parameters affect the total amount of such correlations, highlighting in particular their dependence on the inflationary energy scales and the number of e-foldings during slow-roll.
\end{abstract}


\maketitle

\section{Introduction}

According to the cosmic inflation paradigm \cite{PhysRevD.23.347, LINDE1983177,Riotto:2002yw, RevModPhys.78.537, Baumann:2009ds}, the large-scale structure of the universe can be traced back to the primordial quantum fluctuations associated with one or more inflaton fields. Such fluctuations were then stretched by the accelerated inflationary expansion, so that modes of cosmological interest today crossed the Hubble horizon during inflation and only re-entered at latest stages \cite{Peebles:1980yev, Mukhanov:1990me, Brandenberger:2003vk}, producing, in particular, the temperature and polarization anisotropies observed in the cosmic microwave background (CMB) radiation \cite{Kovac:2002fg, Abazajian:2013vfg, BICEP2:2014owc,Durrer:2015lza,Planck:2019nip}.

Currently, the peculiar features of the CMB radiation are investigated via classical statistical methods \cite{Tristram:2007zz,Planck:2019evm}. Accordingly, we would like to understand the mechanism by which initial quantum fluctuations evolved into classical perturbations or, alternatively, quantify possible quantum signatures in the CMB, even if still undetectable \cite{PhysRevLett.69.3606,PhysRevD.48.2443,PhysRevD.49.788,Lesgourgues:1996jc,Kiefer:1998qe,Martineau:2006ki,Kiefer:2008ku,Martin:2012pea,Das:2013qwa,Burgess:2014eoa,Nelson:2016kjm,Gong:2019yyz}. Within this picture, \emph{quantum entanglement} has emerged as a fundamental tool to characterize the properties of primordial fluctuations and the details of their quantum-to-classical transition \cite{PhysRevD.88.104003,Kukita:2017tpa,PhysRevD.102.043529,Brahma:2021mng,PhysRevD.105.123523,PhysRevD.107.103512,PhysRevD.109.123520,Belfiglio:2025cst}.

When dealing with cosmological perturbations, it is natural to work in momentum space, as the properties of individual perturbation modes are directly related to the length-scales probed by cosmological observations. This correspondence allows to reconstruct the power spectrum and higher order correlation functions, thus providing a direct link with theoretical models of inflation and their predictions. Momentum-space entanglement techniques have been recently introduced in the context of interacting quantum field theories \cite{PhysRevD.86.045014,Kumar:2017ctm,Grignani:2016igg}, and later generalized to inflationary perturbations in Ref. \cite{PhysRevD.102.043529}, where the von Neumann entropy has been employed to compute entanglement between the physically relevant super-Hubble inflationary modes and the remaining \enquote{bath} of sub-Hubble modes. In particular, assuming single-field inflation, it has been shown that the entropy arising from cubic non-Gaussian gravitational interactions \cite{Maldacena:2002vr,Acquaviva:2002ud,Bartolo:2004if} across Hubble horizon is typically much larger than the widely studied squeezing contribution \cite{PhysRevD.50.4807,Gasperini:1992xv,Gasperini:1993mq} emerging from the background accelerated expansion. The effects of short-wavelength modes on observable CMB scales can be also investigated via open quantum system approaches, in the attempt to understand how entanglement may result in observable quantum signatures associated with decoherence processes \cite{Martin:2018zbe,Martin:2018lin,Brahma:2022yxu,Burgess:2022nwu,PhysRevD.108.123530}. If properly singled out, these contributions would inevitably represent a smoking gun for the quantum origin of cosmological perturbations \cite{Lopez:2025arw}.

The above presented investigations typically focus on bipartite entanglement measures, which are defined by appropriately tracing out perturbation modes inside the Hubble horizon during inflation. However, this approach inevitably misses the additional multipartite entanglement features of cubic and higher-order gravitational interactions, which may play a key role in early decoherence processes, starting from reheating\footnote{The possible presence of spectator fields during inflation would speed up decoherence effects, see e.g. \cite{Colas:2022kfu,Colas:2024ysu,Lopez:2025arw} for further discussions.}, when the inflaton field is expected to couple with standard model fields. In particular, the scales currently probed in observations were in super-Hubble form at the end of inflation, thus not being involved in microphysical processes taking place immediately after the slow-roll phase. Accordingly, in order to fully address decoherence processes of inflationary perturbations and their quantum-to-classical transition, it is necessary to describe first their multipartite entanglement features during inflation.

Motivated by this fact, we here employ the recently-proposed Entanglement Distance (ED) \cite{PhysRevA.101.042129} to study additional entanglement features associated with cubic non-Gaussian gravitational interactions in single-field inflationary scenarios. The ED arises from an adapted application of the Fubini-Study metric and, in discrete-variable frameworks, it satisfies all the properties of an entanglement measure, with relevant implications in quantum key distribution protocols for both multi-qubit and multi-qudit quantum systems. A possible characterization of the ED in gravitational particle production processes \cite{PhysRevLett.21.562,PhysRevD.39.389,Ford:2021syk,RevModPhys.96.045005} has been recently investigated within inflationary settings \cite{Belfiglio:2025ofg}. Furthermore, a generalization to continuous variable systems can be derived by studying the Fubini-Study distance between a given multipartite quantum state and the manifold of separable products of coherent states \cite{Vesperini:2023wks}. In order to derive the appropriate multipartite state for inflationary fluctuations, we first quantize the gauge-invariant comoving curvature perturbation and select the Bunch-Davies vacuum as initial state for perturbation modes at the inflationary onset. Within the interaction picture, the squeezing associated with spacetime expansion is naturally encoded in the Bogoliubov transformations for ladder operators. We then introduce non-Gaussian gravitational interactions and compute the corresponding transition amplitudes, focusing on the dominant third-order term for superhorizon processes. Accordingly, the final state for perturbation modes exhibits multipartite entanglement features that cannot be captured by standard bipartite approaches. We show that the ED associated with such a multipartite state is proportional to the number of excitations in the final state, providing a direct physical interpretation of the geometric Fubini-Study distance in the perturbative limit.  Despite not capturing only genuine quantum entanglement, the ED then establishes an upper bound on the von Neumann entropy of any reduced state for perturbations, independently from the choice of modes to be traced out. This is shown by explicitly computing the thermal von Neumann entropy associated with the perturbative number density. Furthermore, we observe that the contribution arising from third-order interactions at the end of the slow-roll regime is typically much larger than the usual squeezing term, in agreement with previous findings on von Neumann entropy generation across the Hubble horizon. We study the dependence of such quantum correlations on the inflationary energy scales and the total duration of the slow-roll regime, highlighting how infrared and ultraviolet cutoffs naturally emerge for momentum modes. We further discuss possible generalizations of the ED for continuous variable systems, with the aim of quantifying genuine multi-mode entanglement associated with inflationary perturbations.

The paper is organized as follows. In Sec. \ref{sezione2}, single-field inflation is reviewed and cubic gravitational interactions are introduced in the dynamics of the comoving curvature perturbation. The corresponding entanglement between perturbation modes is quantified in Sec. \ref{sezione3}, where our measure is employed to derive an upper bound on the von Neumann entropy during slow-roll. Physical consequences are then explored in Sec. \ref{sezione4}, where we also draw our conclusions and present some future perspectives.


\section{Inflationary setup}\label{sezione2}

We consider a scalar inflaton field $\phi$, with corresponding Lagrangian density
\be \label{inf_densl}
\mathcal{L}= \frac{1}{2}  g^{\mu \nu} \phi_{, \mu} \phi_{,\nu}- V(\phi),
\ee
where the potential $V(\phi)$ drives the inflationary phase and $g_{\mu \nu}$ denotes the metric tensor. The dynamics of the inflaton field is typically studied via the standard ansatz \cite{Brandenberger:2003vk}
\be \label{infans}
\phi({\bf x},\tau)=\phi_B(\tau)+ \delta \phi ({\bf x},\tau),
\ee
which separates the homogeneous background contribution, $\phi_B$, from its quantum fluctuations, $\delta \phi$, depending on the position and conformal time, $\tau= \int dt/a(t)$, where $t$ denotes the measurable cosmic time.

The presence of fluctuations induces perturbations on the background spacetime expansion, i.e., $
g_{\mu \nu}= a^2(\tau) \left( \eta_{\mu \nu}+ h_{\mu \nu} \right)$ and $\lvert h_{\mu \nu} \rvert \ll 1$, where $a(\tau)$ is the scale factor and $\eta_{\mu \nu}$ the Minkowski metric tensor. 

We describe the slow-roll of the inflaton field as a quasi-de Sitter phase, selecting \cite{Riotto:2002yw,PhysRevD.101.083516}
\be \label{quasids}
a(\tau)= -\frac{1}{H_I \left(\tau-2\tau_f\right)^{1+\epsilon}},
\ee
where $\tau_f$ denotes the end of the slow-roll phase. Furthermore, we assume a constant Hubble parameter $H_I$, whose value is fixed at horizon crossing for the standard pivot scale, $k_{\rm piv}=0.002$ Mpc$^{-1}$, compatible with the Planck mission constraint \cite{Planck:2018jri}
\be \label{hubble_inf}
H_I < 2.5 \times 10^{-5} \bar{M}_{\rm pl} \simeq 6.1 \times 10^{13}\text{ GeV},
\ee
where $\bar{M}_{\rm pl}$ is the reduced Planck mass. The corresponding slow-roll parameter, $\epsilon$, can be quantified via \cite{Baumann:2009ds,Planck:2018jri},
\be \label{scapow}
\epsilon= \frac{1}{8\pi^2 P_s} \left( \frac{H_I}{\bar{M}_{\rm pl}}  \right)^2,
\ee
denoting by $P_s$ the dimensionless scalar power spectrum, observationally constrained at $P_s=2.1 \times 10^{-9}$ for $k_{\rm piv}$. 

Within single-field inflation, scalar perturbations can be described by a single perturbation potential. Selecting the comoving gauge \cite{PhysRevD.31.1792}, we can write
\be \label{comgau_met}
ds^2= a^2(\tau) \left[ d\tau^2- (1+ 2\zeta) d{\bf x}^2   \right],
\ee
with $\zeta({\bf x},\tau)$ the comoving curvature perturbation. The action for cosmological perturbations has a canonical kinetic term if we appropriately rescale the curvature perturbation via
\be \label{comov_resc}
\chi({\bf x},\tau) = z(\tau) \zeta({\bf x},\tau),
\ee
where $z^2=2 \epsilon a^2 M_{\rm pl}^2$ for standard single-field inflation. Accordingly, $\chi$ can be quantized as 
\be \label{Msresc_exp}
\hat{\chi}({\bf x},\tau)= \frac{1}{\left( 2\pi \right)^3} \int d^3k \left[   \chi_k(\tau) e^{i {\bf k} \cdot {\bf x}} a_{\bf k} + \chi_k^*(\tau) e^{-i {\bf k} \cdot {\bf x}} a^\dagger_{\bf k} \right],
\ee
where the mode functions $v_k(\tau)$ obey the Mukhanov-Sasaki equation
\be 
\chi_k^{\prime \prime} + \left( k^2 - \frac{z^{\prime \prime}}{z}  \right) \chi_k = 0,
\ee
with the prime denoting derivative with respect to conformal time. From Eq. \eqref{quasids}, we obtain 
\be \label{modeq_t}
\chi_k^{\prime \prime}+ \left( k^2- \frac{(2+3\epsilon)}{\eta^2}  \right) \chi_k=0,
\ee
having introduced the rescaled time $\eta \equiv \tau-2\tau_f$. This equation can be solved in terms of Hankel functions, leading to
\be \label{hank_mod} 
\chi_k= \sqrt{-\eta} \left[ c_1 (k) H^{(1)}_\nu(-k\eta)  + c_2(k) H^{(2)}_\nu (-k\eta)  \right],
\ee
with $\nu= \sqrt{9/4+3\epsilon}$, while $c_1(k)$, $c_2(k)$ can be derived by imposing the Bunch-Davies vacuum initial conditions \cite{Bunch:1978yq,Danielsson:2003wb,Greene:2005wk}, giving
\be
\begin{aligned} \label{BD_coe}
& c_1(k)= \frac{\sqrt{\pi}}{2} e^{i \left( \nu + \frac{1}{2} \right) \frac{\pi}{2}}, \\[1pt] 
& c_2(k)=0.
\end{aligned}
\ee
In the limit $\epsilon \ll 1$, Eqs. \eqref{hank_mod}-\eqref{BD_coe} give the simplified expression 
\be \label{simpl_mod}
\chi_k(\eta) \simeq \frac{e^{-ik\eta}}{\sqrt{2k}}\left( 1-\frac{i}{k\eta} \right),
\ee
which is valid within the slow-roll regime, namely $\eta < -\tau_f$. 

\subsection{Cubic non-Gaussianities}

We now focus on the effects of cubic gravitational interactions, which are inevitably present due to the nonlinearity of general relativity. Restricting the analysis to the leading order terms in the slow-roll parameters and ignoring nonlocal contributions, which can be shown to be subdominant in our single-field scenario \cite{PhysRevD.102.043529}, the corresponding action can be expressed as
\be \label{intlag_cub}
S_3= \epsilon^2 M_{\rm pl}^2 \int d\tau d^3x  \left[  \zeta \left( \zeta^\prime \right)^2 + \zeta \left( \partial \zeta   \right)^2  \right] a^2 ,
\ee
where we have assumed a constant slow-roll parameter $\epsilon$, in agreement with Eq. \eqref{quasids}. The above action then contains the dominant terms responsible for entanglement generation between modes. Since perturbation modes $\chi_k$ are frozen out in the limit $k \ll H_c$, where $H_c(\tau)=a(\tau)H_I$, we will focus on the second term in Eq. \eqref{intlag_cub} to compute probability amplitudes, thus defining
\be \label{intlag_dom}
\mathcal{L}_{\rm int}= \epsilon^2 M_{\rm pl}^2 \zeta \left( \partial \zeta \right)^2 a^2.
\ee
Accordingly, working at first order in Dyson expansion, we can write
\begin{align} \label{fin_st}
\ket{\Psi}= &\mathcal{N} \bigg( \ket{0}_{\rm in}+ \frac{1}{6}\int \frac{d^3k_1}{(2\pi)^3} \frac{d^3k_2}{(2\pi)^3} \frac{d^3k_3}{(2\pi)^3} \notag \\
&\ \ \ \ \ \ \ \ \ \ \ \ \ \ \ \  \times \bra{k_1,k_2,k_3} S_{\rm int}  \ket{0}_{\rm in} \ket{k_1,k_2,k_3}  \bigg),
\end{align}
where $S_{\rm int}$ is obtained from $\mathcal{L}_{\rm int}$ and $\ket{0}_{\rm in}$ is the Bunch-Davies initial vacuum state, satisfying $a_{\bf k} \ket{0}_{\rm in}= 0\ \ \forall\ {\bf k}$, while $\mathcal{N}$ is a normalization constant. When computing probability amplitudes, we do not consider modes which are already outside the Hubble horizon at the beginning of inflation. This is equivalent to impose the infrared cutoff $k > a(\tau_i) H_I$, where the initial time $\tau_i$ is determined by selecting a given number of e-foldings before the pivot scale $k_{\rm piv}$ crosses the horizon. From the Planck satellite data, we require
\be \label{efold_ch}
N \gtrsim N_*+ 4.9,
\ee
where $N$ is the total number of inflationary e-foldings and $N_* \equiv \ln \left[ a(\tau_f)/a(\tau_{\rm piv})  \right]$, with $\tau_{\rm piv}$ denoting the time at which the chosen pivot scale crosses the horizon, namely $k_{\rm piv} \equiv a(\tau_{\rm piv}) H_I$. From Eq. \eqref{intlag_dom}, we find
\begin{align}
    \mathcal{C}(k_1,k_2,k_3) & \equiv \bra{k_1,k_2,k_3} S_{\rm int}  \ket{0}_{\rm in} \notag \\
    & =  -i \epsilon^2 (2\pi)^3 M_{\rm pl}^2  \delta\left( {\bf k}_1 + {\bf k}_2+ {\bf k}_3 \right) \notag \\
    & \ \ \ \times (k_1^2+k_2^2+k_3^2) \int_{\tau_i}^{\tau_f} d\tau a^2(\tau)\ \zeta_{k_1}^* \zeta_{k_2}^*  \zeta_{k_3}^*,
\end{align}
where $\zeta_k \equiv \chi_k/z$ is the original perturbation mode


\section{Multipartite entanglement of inflationary perturbations}\label{sezione3}

The presence of cubic and higher order gravitational interactions is responsible for entanglement generation between perturbation modes. In order to quantify the amount of entanglement associated with inflationary perturbations, the Hilbert space of perturbation states is typically divided into two subsystems
\begin{align}
&\mathcal{H}_A(\tau) = \prod \mathcal{H}_k, \ \ \ k < H_c(\tau) \label{supH_space}\\
&\mathcal{H}_B (\tau) = \prod \mathcal{H}_k, \ \ \ k \geq H_c(\tau), \label{subH_space}
\end{align}
where $\mathcal{H}_k$ denotes the harmonic oscillator Hilbert space of the $k$-th mode. The above bipartition naturally allows to compute the von Neumann entropy of a given subsystem with respect to the other \cite{PhysRevD.86.045014,PhysRevD.102.043529}. Since the lengthscales currently probed in CMB observations corresponds to modes crossing the horizon during the early stages of inflation, one typically assumes the space of super-Hubble modes in Eq. \eqref{supH_space} to be the relevant system, with Eq. \eqref{subH_space} representing the bath of sub-Hubble modes\footnote{It must be noted that the separation scale $H_c$, i.e., the inverse of the comoving Hubble radius, is a time-dependent quantity. In particular, the dimension of the system Hilbert space increases with time during inflation.} to be traced out.

\subsection{The Entanglement Distance}

The above approach, while representing an important starting point to address entanglement generation during inflation, inevitably misses the additional multipartite entanglement features induced by cubic gravitational interactions. Specifically, the entanglement between sub-Hubble modes must be taken into account in order to study decoherence processes immediately after inflation, when modes in Eq. \eqref{supH_space} lie outside the Hubble horizon, thus not participating to the microphysical processes taking place at the end of the slow-roll phase. To properly characterize the quantum-to-classical transition of primordial perturbations, we therefore need to derive a general expression for their multipartite entanglement, independent of the separation scale  $H_c$.

Although several theoretical proposals have been developed to quantify multipartite entanglement in quantum information scenarios, \emph{a consistent generalization to relativistic settings is still under investigation.} \cite{Vesperini:2023wks,Belfiglio:2025ofg}. Here, we focus on the recently proposed entanglement distance, an information-geometric measure of entanglement defined through the Fubini-Study metric, which endows the projective Hilbert space of quantum systems with a Riemannian metric structure. For qubit and qudit applications, the ED has been shown to constitute a genuine entanglement measure. However, when moving to continuous variable systems, the situation is typically more complicated, since spanning the full set of local unitary operators which defines equivalence classes of states is an impracticable task. In this context, a proper generalization of the ED can be only formulated for some special classes of states. In particular, for linear combinations of products of coherent states $\ket{s} \in  \otimes_{\mu=0}^n \mathcal{H}_\mu$, with $\mathcal{H}_\mu$ denoting an infinite-dimensional Fock space, the displacement operators
\be \label{dispop}
\left\{ \ket{\mathcal{D}, s}= \prod_{\mu=0}^{M-1} \mathcal{D}^\mu \ket{s} \right\},
\ee
defined by 
\be \label{dispop_sing}
\mathcal{D}^\mu(\alpha^\mu) = \exp \left( \alpha^\mu a^\dagger_\mu - \alpha^{\mu *} a_\mu \right),\ \ \ \ \ \ \alpha^\mu \in \mathbb{C},
\ee
represent an appropriate set of local unitaries. The corresponding ED can be expressed as 
\be \label{ed_coh}
E\left( \ket{s}  \right) = 4 \sum_{\mu=1}^n \left[ \bra{s} a^\dagger_\mu a_\mu \ket{s} - \bra{s} a^\dagger_\mu \ket{s} \bra{s} a_\mu \ket{s} \right],
\ee
thus quantifying the Fubini-Study distance between $\ket{s}$ and the manifold of product coherent states.


\subsection{Multimode inflationary entanglement}

Let us now derive the ED associated with perturbation states during inflation. In order to properly include the squeezing effects related to spacetime evolution, we start by defining the standard Bogoliubov transformation
\be \label{sq_ops}
b_{\bf k} = \alpha_k a_{\bf k} + \beta_k^* a^\dagger_{-{\bf k}},
\ee
where $\alpha_k$ and $\beta_k$ are the Bogoliubov coefficients associated with the background expansion, while $a_{\bf k}$ annihilates the Bunch-Davies vacuum, in agreement with Eq. \eqref{Msresc_exp}. In the de Sitter limit ($\epsilon=0$), they are given by \cite{PhysRevD.50.4807,PhysRevD.102.043529}
\begin{align} \label{}
&\alpha_k = e^{i \theta_k(\eta)} \cosh\left[  r_k(\eta) \right],\\
&\beta_k= e^{-i \theta_k(\eta)+ 2i \phi_k(\eta)} \sinh\left[  r_k(\eta) \right],
\end{align}
where
\begin{align}
    & \theta_k(\eta) =  -k\eta - \tan^{-1}\left( \frac{1}{2k\eta} \right), \\[1.5pt]
    & \phi_k(\eta) = - \frac{\pi}{4}- \frac{1}{2}\tan^{-1}\left( \frac{1}{2k\eta} \right),  \\[1.5pt]
    & r_k(\eta) = -\sinh^{-1} \left( \frac{1}{2k\eta} \right). [1.5pt]
\end{align}
In particular, the parameter $r_k$ quantifies the amount of squeezing associated with each inflationary mode, which increases in time during the slow-roll regime\footnote{This approach has been recently criticized, highlighting that the notion of squeezed states is subject to ambiguities when quantum fields evolve in time-dependent backgrounds and asymptotic flatness is not appropriately recovered \cite{Agullo:2022ttg}. However, we will see that the squeezing contribution associated with the ED is typically negligible.}.

    \begin{figure*}
    \centering
    \includegraphics[scale=1.2]{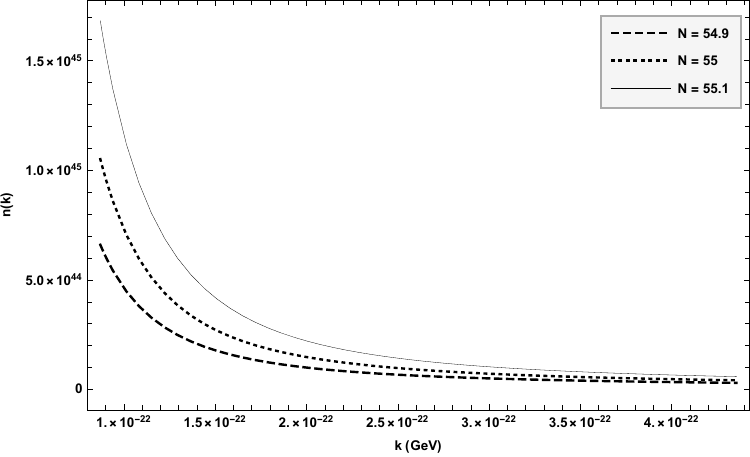}
    \caption{Number density $n(k)$ as a function of the comoving wavenumber $k$, assuming slight variations in the total number of inflationary e-foldings $N$. We select $k \lesssim a(\tau_f) H_I$, further setting $H_I= 4 \times 10^{13}$ GeV and $N-N_*=10$, in agreement with Planck data. }
    \label{fig_nudens}
\end{figure*}

Accordingly, we can write the ED for the state $\ket{\Psi}$ in Eq. \eqref{fin_st} as
\begin{align} \label{ed_pert}
E\left( \ket{\Psi} \right) &= 4 \int \frac{d^3k}{(2\pi)^3} \big[ \bra{\Psi} b_{\bf k}^\dagger b_{\bf k} \ket{\Psi} \notag \\
&\ \ \ \ \ \ \ \ \ \ \ \ \ \ \ \ \ \ \ - \bra{\Psi} b^\dagger_{\bf k} \ket{\Psi} \bra{\Psi} b_{\bf k} \ket{\Psi}  \big],
\end{align}
where, to ensure finite probability amplitudes, we further need to impose the standard ultraviolet cutoff $k < a(\tau_f) M_{\rm pl}$. Since the second term in the above equation is zero, we notice that the ED is here proportional to the expectation value of the number operator, implying
\be \label{numbdens}
V \int d^3k\ n (k) =  \frac{E\left( \ket{\Psi} \right)}{4},
\ee
where $V= (2\pi)^3 \delta\left( {\bf k} - {\bf k} \right)$ is the standard volume contribution \cite{RevModPhys.96.045005} and $n(k)$ the comoving number density. In Fig. \ref{fig_nudens}, we show the number density scaling for different values of the total inflationary e-foldings $N$, selecting field modes which are close to the Hubble horizon at the end of inflation. Rewriting now Eq. \eqref{ed_pert} in the form
\be \label{ed_split}
E\left( \ket{\Psi} \right) = E_{\rm sq} + E_{\rm cub} ,
\ee
where 
\begin{align}
    & E_{\rm sq} = 4 V \lvert \mathcal{N} \rvert^2 \int \frac{d^3k}{\left(2\pi \right)^3} \lvert \beta_k \rvert^2, \\[2.5pt]
    & E_{\rm cub} =  \frac{\lvert \mathcal{N} \rvert^2}{9} \int \frac{d^3k_1}{(2\pi)^3} \frac{d^3k_2}{(2\pi)^3} \frac{d^3k_3}{(2\pi)^3} \lvert \mathcal{C}(k_1,k_2,k_3) \rvert^2 \notag \\
    &\ \ \ \ \ \ \ \ \ \ \ \ \ \ \ \ \ \ \ \ \ \times \left( 3 +  2\sum_{i=1}^3 \lvert \beta_{k_i} \rvert^2 \right),
\end{align}
it can be shown that $E_{\rm sq} \ll E_{\rm cub}$ and, in particular,
\be 
\frac{E_{\rm cub}}{E_{\rm sq}} \simeq 10^9 \left( \frac{H_I}{M_{\rm pl}} \right) e^{2N},
\ee
in agreement with the results of Ref. \cite{PhysRevD.102.043529}, which focused the von Neumann entropy of super-Hubble modes, once the environment of sub-Hubble modes is traced out. 
\begin{table}[t!] 
\centering 
\resizebox{8.1cm}{!}{
\begin{tabular}{|c|c|c|}
\hline\hline
$H_I$ ($10^{13}$ GeV) & $\mathcal{S}_{\rm th}$ ($10^{-47}$ GeV$^{3}$) & $s_{\rm th}$ ($10^{57}$ GeV$^{3}$)\\
\hline\hline
1.0 & 22.9  & 2.641 \\ \hline
1.5 & 6.996  & 2.666 \\ \hline
2.0 & 3.06  & 2.684 \\ \hline
2.5 & 1.636  & 2.699 \\ \hline
3.0 & 0.996  & 2.711 \\ \hline
3.5 & 0.665  & 2.721 \\ \hline
4.0 & 0.476  & 2.731 \\ \hline
\hline
\end{tabular}}
\caption{Thermal entropy density per comoving volume, $\mathcal{S}_{\rm th}$ and physical volume, $s_{\rm th}$, as function of the inflationary Hubble parameter $H_I$. We set $N=55$ and $N_*=45$.}
\label{tab_enta}
\end{table}

The presence of non-Gaussian terms and squeezing effects implies that the ED in Eq. \eqref{ed_pert} is not able to provide a genuine measure of multipartite entanglement in the case of inflationary perturbations. However, once obtained the particle number density in Eq. \eqref{numbdens}, it allows to derive an upper bound on the von Neumann entropy density $\mathcal{S}$ arising from the non-Gaussian interactions, independently of the selected bipartition. At fixed $n (k)$, the entropy density is indeed maximized by thermal states, having 
\be \label{max_ent}
\mathcal{S}_{\rm th} = \int \frac{d^3 k}{(2\pi)^3} \left[  \left( 1+n(k) \right)  \ln \left( 1+n(k) \right) - n(k) \ln n(k)  \right].
\ee
In Tab. \ref{tab_enta}, we display $\mathcal{S}_{\rm th}$ and the corresponding entropy per physical volume element, namely $s_{\rm th} \equiv \mathcal{S}_{\rm th}/a^3(\tau_f)$, by varying the inflationary energy scale via the 
Hubble parameter $H_I$. 

Our outcomes readily implies that, once defined the reduced density operator
\be \label{red_ent}
\rho_A = \text{Tr}_{\bar{A}} \ket{\Psi} \bra{\Psi},
\ee
where $A$ is a subset of comoving momenta, namely $A \subset \left[a(\tau_i) H_I, a(\tau_f) M_{\rm pl} \right]$, then $\mathcal{S}(\rho_A) < \mathcal{S}_{\rm th}$ independently of the choice of the subset.


\section{Final outlooks and perspectives}\label{sezione4}

In this work, we explored the emergence of multipartite quantum correlations during single-field inflation. In particular, we studied the dynamics of the comoving curvature perturbation by focusing on non-Gaussian gravitational interactions, which represent a plausible source of quantum signatures in the CMB radiation. 

Within the interaction picture, we first derived the final multipartite quantum state of perturbation modes, retaining the leading-order cubic contributions in the slow-roll parameters and neglecting non-local terms. Furthermore, we included squeezing effects due to spacetime expansion through standard Bogoliubov transformations. 

By employing the recently-proposed ED, we then provided a geometric and operational characterization scheme of the quantum state of inflationary fluctuations. In particular, we computed the dominant third-order transition amplitudes, showing that the ED associated with cubic interactions scales proportionally to the total number of excitations generated in the final state. 

Accordingly, our findings offer a physical interpretation of the Fubini-Study distance within the perturbative regime and establish an upper bound on the von Neumann entropy density of any reduced perturbation state compatible with such excitations. We computed the corresponding entropy bound as a thermal contribution, which is then independent of the choice of traced-out modes.

In addition, our outcomes revealed that the quantum correlations generated by cubic non-Gaussian interactions typically dominate over the standard contribution from squeezing, confirming and extending previous results on entropy generation across the Hubble horizon, see e.g. \cite{PhysRevD.102.043529, PhysRevD.109.123520}. Particularly, we found that the magnitude of such quantum correlations is significantly influenced by the total duration of the slow-roll phase and the inflationary energy scales, with both infrared and ultraviolet cutoffs naturally emerging from the momentum-domain structure of the interaction integrals.

Finally, we discussed how the ED requires further generalization in order to represent a genuine multi-mode entanglement quantifier directly applicable to cosmological perturbations. In particular, the inclusion of squeezing and non-Gaussian correlations would affect the Fubini-Study metric construction, and thus the corresponding distance between quantum states. These developments could, in turn, provide new tools to connect inflationary dynamics with observational probes of primordial non-Gaussianity, especially once post-inflationary decoherence effects are properly included. 

Future works will clarify how to refine our underlying entanglement measure and how to identify possible observational signatures associated with it. In the era of precision cosmology, understanding how to measure entanglement may represent a key step to address the quantum-to-classical transition of cosmological perturbations and to probe the quantum nature of gravity, still a subject of profound speculation. Further, we intend to study possible connections between the hypothesis of emergent gravity and our approach, seeking plausible intersections between quantum information world and cosmology, see e.g. \cite{Verlinde:2016toy,Faulkner:2017tkh}.

\section*{Acknowledgements} 
The authors express their gratitude to Stefano Mancini for intriguing discussions on topics related to this work. AB and RF would like to acknowledge INFN Pisa for the financial support to this activity. RF acknowledges the support of the Research Support Plan 2022 - Call for applications for funding allocation to research projects curiosity-driven (F CUR) - Project \enquote{Entanglement Protection of Qubits' Dynamics in a Cavity} - EPQDC. OL acknowledges support by the  Fondazione  ICSC, Spoke 3 Astrophysics and Cosmos Observations. National Recovery and Resilience Plan (Piano Nazionale di Ripresa e Resilienza, PNRR) Project ID $CN00000013$ ``Italian Research Center on  High-Performance Computing, Big Data and Quantum Computing" funded by MUR Missione 4 Componente 2 Investimento 1.4: Potenziamento strutture di ricerca e creazione di ``campioni nazionali di R\&S (M4C2-19)" - Next Generation EU (NGEU). AB and OL also grateful to Leonardo Rossetti for discussions on the Fubini-Study metric and possible connections in the phase-space of the primordial universe.

\bibliographystyle{unsrt}
\bibliography{bibliog}

@article{Riotto:2002yw,
    author = "Riotto, Antonio",
    editor = "Dvali, G. and Perez-Lorenzana, Abdel and Senjanovic, G. and Thompson, G. and Vissani, F.",
    title = "{Inflation and the theory of cosmological perturbations}",
    eprint = "hep-ph/0210162",
    archivePrefix = "arXiv",
    reportNumber = "DFPD-TH-02-22",
    journal = "ICTP Lect. Notes Ser.",
    volume = "14",
    pages = "317--413",
    year = "2003"
}

@article{RevModPhys.78.537,
  title = {Inflation dynamics and reheating},
  author = {Bassett, Bruce A. and Tsujikawa, Shinji and Wands, David},
  journal = {Rev. Mod. Phys.},
  volume = {78},
  issue = {2},
  pages = {537--589},
  numpages = {0},
  year = {2006},
  month = {May},
  publisher = {American Physical Society},
  doi = {10.1103/RevModPhys.78.537},
  url = {https://link.aps.org/doi/10.1103/RevModPhys.78.537}
}

@inproceedings{Baumann:2009ds,
    author = "Baumann, Daniel",
    title = "{Inflation}",
    booktitle = "{Theoretical Advanced Study Institute in Elementary Particle Physics}: {Physics of the Large and the Small}",
    eprint = "0907.5424",
    archivePrefix = "arXiv",
    primaryClass = "hep-th",
    reportNumber = "TASI-2009",
    doi = "10.1142/9789814327183_0010",
    pages = "523--686",
    year = "2011"
}

@book{Peebles:1980yev,
    author = "Peebles, P. James",
    title = "{The Large-Scale Structure of the Universe}",
    isbn = "978-0-691-08240-0, 978-0-691-20983-8, 978-0-691-20671-4",
    publisher = "Princeton University Press",
    month = "11",
    year = "1980"
}

@article{Mukhanov:1990me,
    author = "Mukhanov, Viatcheslav F. and Feldman, H. A. and Brandenberger, Robert H.",
    title = "{Theory of cosmological perturbations. Part 1. Classical perturbations. Part 2. Quantum theory of perturbations. Part 3. Extensions}",
    reportNumber = "BROWN-HET-796, BROWN-HET-800, BROWN-HET-780",
    doi = "10.1016/0370-1573(92)90044-Z",
    journal = "Phys. Rept.",
    volume = "215",
    pages = "203--333",
    year = "1992"
}

@article{Brandenberger:2003vk,
    author = "Brandenberger, Robert H.",
    editor = "Breton, Nora and Cervantes-Cota, Jorge L. and Salgado, Marcelo",
    title = "{Lectures on the theory of cosmological perturbations}",
    eprint = "hep-th/0306071",
    archivePrefix = "arXiv",
    reportNumber = "BROWN-HET-1358",
    doi = "10.1007/978-3-540-40918-2_5",
    journal = "Lect. Notes Phys.",
    volume = "646",
    pages = "127--167",
    year = "2004"
}

@article{Martin:2018zbe,
    author = "Martin, Jerome and Vennin, Vincent",
    title = "{Observational constraints on quantum decoherence during inflation}",
    eprint = "1801.09949",
    archivePrefix = "arXiv",
    primaryClass = "astro-ph.CO",
    doi = "10.1088/1475-7516/2018/05/063",
    journal = "JCAP",
    volume = "05",
    pages = "063",
    year = "2018"
}

@article{PhysRevD.101.083516,
  title = {Nonadiabatic cosmological production of ultralight dark matter},
  author = {Herring, Nathan and Boyanovsky, Daniel and Zentner, Andrew R.},
  journal = {Phys. Rev. D},
  volume = {101},
  issue = {8},
  pages = {083516},
  numpages = {23},
  year = {2020},
  month = {Apr},
  publisher = {American Physical Society},
  doi = {10.1103/PhysRevD.101.083516},
  url = {https://link.aps.org/doi/10.1103/PhysRevD.101.083516}
}

@article{Martin:2018lin,
    author = "Martin, J{\'e}r{\^o}me and Vennin, Vincent",
    title = "{Non Gaussianities from Quantum Decoherence during Inflation}",
    eprint = "1805.05609",
    archivePrefix = "arXiv",
    primaryClass = "astro-ph.CO",
    doi = "10.1088/1475-7516/2018/06/037",
    journal = "JCAP",
    volume = "06",
    pages = "037",
    year = "2018"
}

@article{Kovac:2002fg,
    author = "Kovac, John and Leitch, E. M. and Pryke, C and Carlstrom, J. E. and Halverson, N. W. and Holzapfel, W. L.",
    title = "{Detection of polarization in the cosmic microwave background using DASI}",
    eprint = "astro-ph/0209478",
    archivePrefix = "arXiv",
    doi = "10.1038/nature01269",
    journal = "Nature",
    volume = "420",
    pages = "772--787",
    year = "2002"
}

@article{Abazajian:2013vfg,
    author = "Abazajian, K. N. and others",
    title = "{Inflation Physics from the Cosmic Microwave Background and Large Scale Structure}",
    eprint = "1309.5381",
    archivePrefix = "arXiv",
    primaryClass = "astro-ph.CO",
    reportNumber = "FERMILAB-PUB-13-442-A",
    doi = "10.1016/j.astropartphys.2014.05.013",
    journal = "Astropart. Phys.",
    volume = "63",
    pages = "55--65",
    year = "2015"
}

@article{BICEP2:2014owc,
    author = "Ade, P. A. R. and others",
    collaboration = "BICEP2",
    title = "{Detection of $B$-Mode Polarization at Degree Angular Scales by BICEP2}",
    eprint = "1403.3985",
    archivePrefix = "arXiv",
    primaryClass = "astro-ph.CO",
    doi = "10.1103/PhysRevLett.112.241101",
    journal = "Phys. Rev. Lett.",
    volume = "112",
    number = "24",
    pages = "241101",
    year = "2014"
}

@article{Planck:2019nip,
    author = "Aghanim, N. and others",
    collaboration = "Planck",
    title = "{Planck 2018 results. V. CMB power spectra and likelihoods}",
    eprint = "1907.12875",
    archivePrefix = "arXiv",
    primaryClass = "astro-ph.CO",
    doi = "10.1051/0004-6361/201936386",
    journal = "Astron. Astrophys.",
    volume = "641",
    pages = "A5",
    year = "2020"
}

@article{Planck:2018jri,
    author = "Akrami, Y. and others",
    collaboration = "Planck",
    title = "{Planck 2018 results. X. Constraints on inflation}",
    eprint = "1807.06211",
    archivePrefix = "arXiv",
    primaryClass = "astro-ph.CO",
    doi = "10.1051/0004-6361/201833887",
    journal = "Astron. Astrophys.",
    volume = "641",
    pages = "A10",
    year = "2020"
}

@article{Maldacena:2002vr,
    author = "Maldacena, Juan Martin",
    title = "{Non-Gaussian features of primordial fluctuations in single field inflationary models}",
    eprint = "astro-ph/0210603",
    archivePrefix = "arXiv",
    doi = "10.1088/1126-6708/2003/05/013",
    journal = "JHEP",
    volume = "05",
    pages = "013",
    year = "2003"
}

@article{Acquaviva:2002ud,
    author = "Acquaviva, Viviana and Bartolo, Nicola and Matarrese, Sabino and Riotto, Antonio",
    title = "{Second order cosmological perturbations from inflation}",
    eprint = "astro-ph/0209156",
    archivePrefix = "arXiv",
    reportNumber = "DFPD-A-02-21",
    doi = "10.1016/S0550-3213(03)00550-9",
    journal = "Nucl. Phys. B",
    volume = "667",
    pages = "119--148",
    year = "2003"
}

@article{Bartolo:2004if,
    author = "Bartolo, N. and Komatsu, E. and Matarrese, Sabino and Riotto, A.",
    title = "{Non-Gaussianity from inflation: Theory and observations}",
    eprint = "astro-ph/0406398",
    archivePrefix = "arXiv",
    reportNumber = "DFPD-04-A-12",
    doi = "10.1016/j.physrep.2004.08.022",
    journal = "Phys. Rept.",
    volume = "402",
    pages = "103--266",
    year = "2004"
}

@article{PhysRevLett.69.3606,
  title = {Entropy of a classical stochastic field and cosmological perturbations},
  author = {Brandenberger, R. and Mukhanov, V. and Prokopec, T.},
  journal = {Phys. Rev. Lett.},
  volume = {69},
  issue = {25},
  pages = {3606--3609},
  numpages = {0},
  year = {1992},
  month = {Dec},
  publisher = {American Physical Society},
  doi = {10.1103/PhysRevLett.69.3606},
  url = {https://link.aps.org/doi/10.1103/PhysRevLett.69.3606}
}

@article{PhysRevD.49.788,
  title = {Coherent state representation of quantum fluctuations in the early Universe},
  author = {Matacz, A. L.},
  journal = {Phys. Rev. D},
  volume = {49},
  issue = {2},
  pages = {788--798},
  numpages = {0},
  year = {1994},
  month = {Jan},
  publisher = {American Physical Society},
  doi = {10.1103/PhysRevD.49.788},
  url = {https://link.aps.org/doi/10.1103/PhysRevD.49.788}
}

@article{PhysRevD.48.2443,
  title = {Entropy of the gravitational field},
  author = {Brandenberger, R. and Mukhanov, V. and Prokopec, T.},
  journal = {Phys. Rev. D},
  volume = {48},
  issue = {6},
  pages = {2443--2455},
  numpages = {0},
  year = {1993},
  month = {Sep},
  publisher = {American Physical Society},
  doi = {10.1103/PhysRevD.48.2443},
  url = {https://link.aps.org/doi/10.1103/PhysRevD.48.2443}
}

@article{PhysRevD.23.347,
  title = {Inflationary universe: A possible solution to the horizon and flatness problems},
  author = {Guth, Alan H.},
  journal = {Phys. Rev. D},
  volume = {23},
  issue = {2},
  pages = {347--356},
  numpages = {0},
  year = {1981},
  month = {Jan},
  publisher = {American Physical Society},
  doi = {10.1103/PhysRevD.23.347},
  url = {https://link.aps.org/doi/10.1103/PhysRevD.23.347}
}

@article{Gasperini:1992xv,
    author = "Gasperini, M. and Giovannini, Massimo",
    title = "{Entropy production in the cosmological amplification of the vacuum fluctuations}",
    eprint = "gr-qc/9301010",
    archivePrefix = "arXiv",
    reportNumber = "DFTT-63-92",
    doi = "10.1016/0370-2693(93)91159-K",
    journal = "Phys. Lett. B",
    volume = "301",
    pages = "334--338",
    year = "1993"
}

@article{Gasperini:1993mq,
    author = "Gasperini, M. and Giovannini, Massimo",
    title = "{Quantum squeezing and cosmological entropy production}",
    eprint = "gr-qc/9307024",
    archivePrefix = "arXiv",
    reportNumber = "CERN-TH-6954-93",
    doi = "10.1088/0264-9381/10/9/004",
    journal = "Class. Quant. Grav.",
    volume = "10",
    pages = "L133--L136",
    year = "1993"
}

@article{LINDE1983177,
title = {Chaotic inflation},
journal = {Physics Letters B},
volume = {129},
number = {3},
pages = {177-181},
year = {1983},
issn = {0370-2693},
doi = {https://doi.org/10.1016/0370-2693(83)90837-7},
url = {https://www.sciencedirect.com/science/article/pii/0370269383908377},
author = {A.D. Linde},
}

@article{Planck:2019evm,
    author = "Akrami, Y. and others",
    collaboration = "Planck",
    title = "{Planck 2018 results. VII. Isotropy and Statistics of the CMB}",
    eprint = "1906.02552",
    archivePrefix = "arXiv",
    primaryClass = "astro-ph.CO",
    doi = "10.1051/0004-6361/201935201",
    journal = "Astron. Astrophys.",
    volume = "641",
    pages = "A7",
    year = "2020"
}

@article{Brahma:2021mng,
    author = "Brahma, Suddhasattwa and Berera, Arjun and Calder{\'o}n-Figueroa, Jaime",
    title = "{Universal signature of quantum entanglement across cosmological distances}",
    eprint = "2107.06910",
    archivePrefix = "arXiv",
    primaryClass = "hep-th",
    doi = "10.1088/1361-6382/aca066",
    journal = "Class. Quant. Grav.",
    volume = "39",
    number = "24",
    pages = "245002",
    year = "2022"
}

@article{Brahma:2022yxu,
    author = "Brahma, Suddhasattwa and Berera, Arjun and Calder{\'o}n-Figueroa, Jaime",
    title = "{Quantum corrections to the primordial tensor spectrum: open EFTs {\&} Markovian decoupling of UV modes}",
    eprint = "2206.05797",
    archivePrefix = "arXiv",
    primaryClass = "hep-th",
    doi = "10.1007/JHEP08(2022)225",
    journal = "JHEP",
    volume = "08",
    pages = "225",
    year = "2022"
}

@article{Burgess:2022nwu,
    author = "Burgess, C. P. and Holman, R. and Kaplanek, Greg and Martin, Jerome and Vennin, Vincent",
    title = "{Minimal decoherence from inflation}",
    eprint = "2211.11046",
    archivePrefix = "arXiv",
    primaryClass = "hep-th",
    reportNumber = "CERN-TH-2022-174; Imperial/TP/2022/GK/02",
    doi = "10.1088/1475-7516/2023/07/022",
    journal = "JCAP",
    volume = "07",
    pages = "022",
    year = "2023"
}

@article{PhysRevD.108.123530,
  title = {Quantum thermodynamics of de Sitter space},
  author = {Alicki, Robert and Barenboim, Gabriela and Jenkins, Alejandro},
  journal = {Phys. Rev. D},
  volume = {108},
  issue = {12},
  pages = {123530},
  numpages = {13},
  year = {2023},
  month = {Dec},
  publisher = {American Physical Society},
  doi = {10.1103/PhysRevD.108.123530},
  url = {https://link.aps.org/doi/10.1103/PhysRevD.108.123530}
}

@article{Colas:2022kfu,
    author = "Colas, Thomas and Grain, Julien and Vennin, Vincent",
    title = "{Quantum recoherence in the early universe}",
    eprint = "2212.09486",
    archivePrefix = "arXiv",
    primaryClass = "gr-qc",
    doi = "10.1209/0295-5075/acdd94",
    journal = "EPL",
    volume = "142",
    number = "6",
    pages = "69002",
    year = "2023"
}

@article{Colas:2024ysu,
    author = "Colas, Thomas and Grain, Julien and Kaplanek, Greg and Vennin, Vincent",
    title = "{In-in formalism for the entropy of quantum fields in curved spacetimes}",
    eprint = "2406.17856",
    archivePrefix = "arXiv",
    primaryClass = "hep-th",
    doi = "10.1088/1475-7516/2024/08/047",
    journal = "JCAP",
    volume = "08",
    pages = "047",
    year = "2024"
}

@article{Lopez:2025arw,
    author = "Lopez, Francescopaolo and Bartolo, Nicola",
    title = "{Quantum signatures and decoherence during inflation from deep subhorizon perturbations}",
    eprint = "2503.23150",
    archivePrefix = "arXiv",
    primaryClass = "astro-ph.CO",
    month = "3",
    year = "2025"
}

@article{Durrer:2015lza,
    author = "Durrer, Ruth",
    title = "{The cosmic microwave background: the history of its experimental investigation and its significance for cosmology}",
    eprint = "1506.01907",
    archivePrefix = "arXiv",
    primaryClass = "astro-ph.CO",
    doi = "10.1088/0264-9381/32/12/124007",
    journal = "Class. Quant. Grav.",
    volume = "32",
    number = "12",
    pages = "124007",
    year = "2015"
}

@article{Tristram:2007zz,
    author = "Tristram, Matthieu and Ganga, Ken",
    title = "{Data analysis methods for the cosmic microwave background}",
    eprint = "0708.1429",
    archivePrefix = "arXiv",
    primaryClass = "astro-ph",
    doi = "10.1088/0034-4885/70/6/R02",
    journal = "Rept. Prog. Phys.",
    volume = "70",
    pages = "899",
    year = "2007"
}

@article{PhysRevD.88.104003,
  title = {Quantum entanglement generation in de Sitter spacetime},
  author = {Hu, Jiawei and Yu, Hongwei},
  journal = {Phys. Rev. D},
  volume = {88},
  issue = {10},
  pages = {104003},
  numpages = {5},
  year = {2013},
  month = {Nov},
  publisher = {American Physical Society},
  doi = {10.1103/PhysRevD.88.104003},
  url = {https://link.aps.org/doi/10.1103/PhysRevD.88.104003}
}

@article{Kukita:2017tpa,
    author = "Kukita, Shingo and Nambu, Yasusada",
    title = "{Entanglement dynamics in de Sitter spacetime}",
    eprint = "1706.09175",
    archivePrefix = "arXiv",
    primaryClass = "gr-qc",
    doi = "10.1088/1361-6382/aa8e31",
    journal = "Class. Quant. Grav.",
    volume = "34",
    number = "23",
    pages = "235010",
    year = "2017"
}

@article{PhysRevD.102.043529,
  title = {Entanglement entropy of cosmological perturbations},
  author = {Brahma, Suddhasattwa and Alaryani, Omar and Brandenberger, Robert},
  journal = {Phys. Rev. D},
  volume = {102},
  issue = {4},
  pages = {043529},
  numpages = {20},
  year = {2020},
  month = {Aug},
  publisher = {American Physical Society},
  doi = {10.1103/PhysRevD.102.043529},
  url = {https://link.aps.org/doi/10.1103/PhysRevD.102.043529}
}

@article{PhysRevD.50.4807,
  title = {Inflation and squeezed quantum states},
  author = {Albrecht, Andreas and Ferreira, Pedro and Joyce, Michael and Prokopec, Tomislav},
  journal = {Phys. Rev. D},
  volume = {50},
  issue = {8},
  pages = {4807--4820},
  numpages = {0},
  year = {1994},
  month = {Oct},
  publisher = {American Physical Society},
  doi = {10.1103/PhysRevD.50.4807},
  url = {https://link.aps.org/doi/10.1103/PhysRevD.50.4807}
}

@article{Lesgourgues:1996jc,
    author = "Lesgourgues, Julien and Polarski, David and Starobinsky, Alexei A.",
    title = "{Quantum to classical transition of cosmological perturbations for nonvacuum initial states}",
    eprint = "gr-qc/9611019",
    archivePrefix = "arXiv",
    reportNumber = "LMPT-10-96, LMPT 10/96",
    doi = "10.1016/S0550-3213(97)00224-1",
    journal = "Nucl. Phys. B",
    volume = "497",
    pages = "479--510",
    year = "1997"
}

@article{Kiefer:2008ku,
    author = "Kiefer, Claus and Polarski, David",
    title = "{Why do cosmological perturbations look classical to us?}",
    eprint = "0810.0087",
    archivePrefix = "arXiv",
    primaryClass = "astro-ph",
    doi = "10.1166/asl.2009.1023",
    journal = "Adv. Sci. Lett.",
    volume = "2",
    pages = "164--173",
    year = "2009"
}

@article{Kiefer:1998qe,
    author = "Kiefer, Claus and Polarski, David and Starobinsky, Alexei A.",
    title = "{Quantum to classical transition for fluctuations in the early universe}",
    eprint = "gr-qc/9802003",
    archivePrefix = "arXiv",
    reportNumber = "THEP-97-33, FREIBURG-THEP-97-33, Freiburg THEP-97/33",
    doi = "10.1142/S0218271898000292",
    journal = "Int. J. Mod. Phys. D",
    volume = "7",
    pages = "455--462",
    year = "1998"
}

@article{Martineau:2006ki,
    author = "Martineau, Patrick",
    title = "{On the decoherence of primordial fluctuations during inflation}",
    eprint = "astro-ph/0601134",
    archivePrefix = "arXiv",
    doi = "10.1088/0264-9381/24/23/006",
    journal = "Class. Quant. Grav.",
    volume = "24",
    pages = "5817--5834",
    year = "2007"
}

@article{Martin:2012pea,
    author = "Martin, Jerome and Vennin, Vincent and Peter, Patrick",
    title = "{Cosmological Inflation and the Quantum Measurement Problem}",
    eprint = "1207.2086",
    archivePrefix = "arXiv",
    primaryClass = "hep-th",
    doi = "10.1103/PhysRevD.86.103524",
    journal = "Phys. Rev. D",
    volume = "86",
    pages = "103524",
    year = "2012"
}

@article{Das:2013qwa,
    author = "Das, Suratna and Lochan, Kinjalk and Sahu, Satyabrata and Singh, T. P.",
    title = "{Quantum to classical transition of inflationary perturbations: Continuous spontaneous localization as a possible mechanism}",
    eprint = "1304.5094",
    archivePrefix = "arXiv",
    primaryClass = "astro-ph.CO",
    doi = "10.1103/PhysRevD.88.085020",
    journal = "Phys. Rev. D",
    volume = "88",
    number = "8",
    pages = "085020",
    year = "2013",
    note = "[Erratum: Phys.Rev.D 89, 109902 (2014)]"
}

@article{Burgess:2014eoa,
    author = "Burgess, C. P. and Holman, R. and Tasinato, G. and Williams, M.",
    title = "{EFT Beyond the Horizon: Stochastic Inflation and How Primordial Quantum Fluctuations Go Classical}",
    eprint = "1408.5002",
    archivePrefix = "arXiv",
    primaryClass = "hep-th",
    reportNumber = "CERN-PH-TH-2014-142",
    doi = "10.1007/JHEP03(2015)090",
    journal = "JHEP",
    volume = "03",
    pages = "090",
    year = "2015"
}

@article{Nelson:2016kjm,
    author = "Nelson, Elliot",
    title = "{Quantum Decoherence During Inflation from Gravitational Nonlinearities}",
    eprint = "1601.03734",
    archivePrefix = "arXiv",
    primaryClass = "gr-qc",
    doi = "10.1088/1475-7516/2016/03/022",
    journal = "JCAP",
    volume = "03",
    pages = "022",
    year = "2016"
}

@article{Gong:2019yyz,
    author = "Gong, Jinn-Ouk and Seo, Min-Seok",
    title = "{Quantum non-linear evolution of inflationary tensor perturbations}",
    eprint = "1903.12295",
    archivePrefix = "arXiv",
    primaryClass = "hep-th",
    doi = "10.1007/JHEP05(2019)021",
    journal = "JHEP",
    volume = "05",
    pages = "021",
    year = "2019"
}

@article{Agullo:2022ttg,
    author = "Agullo, Ivan and Bonga, B{\'e}atrice and Metidieri, Patricia Ribes",
    title = "{Does inflation squeeze cosmological perturbations?}",
    eprint = "2203.07066",
    archivePrefix = "arXiv",
    primaryClass = "gr-qc",
    doi = "10.1088/1475-7516/2022/09/032",
    journal = "JCAP",
    volume = "09",
    pages = "032",
    year = "2022"
}

@article{PhysRevD.86.045014,
  title = {Momentum-space entanglement and renormalization in quantum field theory},
  author = {Balasubramanian, Vijay and McDermott, Michael B. and Van Raamsdonk, Mark},
  journal = {Phys. Rev. D},
  volume = {86},
  issue = {4},
  pages = {045014},
  numpages = {18},
  year = {2012},
  month = {Aug},
  publisher = {American Physical Society},
  doi = {10.1103/PhysRevD.86.045014},
  url = {https://link.aps.org/doi/10.1103/PhysRevD.86.045014}
}

@article{Kumar:2017ctm,
    author = "Kumar, S. Santhosh and Shankaranarayanan, S.",
    title = "{Role of spatial higher order derivatives in momentum space entanglement}",
    eprint = "1702.08655",
    archivePrefix = "arXiv",
    primaryClass = "hep-th",
    doi = "10.1103/PhysRevD.95.065023",
    journal = "Phys. Rev. D",
    volume = "95",
    number = "6",
    pages = "065023",
    year = "2017"
}

@article{Grignani:2016igg,
    author = "Grignani, Gianluca and Semenoff, Gordon W.",
    title = "{Scattering and momentum space entanglement}",
    eprint = "1612.08858",
    archivePrefix = "arXiv",
    primaryClass = "hep-th",
    doi = "10.1016/j.physletb.2017.07.030",
    journal = "Phys. Lett. B",
    volume = "772",
    pages = "699--702",
    year = "2017"
}

@article{PhysRevA.101.042129,
  title = {Entanglement distance for arbitrary $M$-qudit hybrid systems},
  author = {Cocchiarella, Denise and Scali, Stefano and Ribisi, Salvatore and Nardi, Bianca and Bel-Hadj-Aissa, Ghofrane and Franzosi, Roberto},
  journal = {Phys. Rev. A},
  volume = {101},
  issue = {4},
  pages = {042129},
  numpages = {9},
  year = {2020},
  month = {Apr},
  publisher = {American Physical Society},
  doi = {10.1103/PhysRevA.101.042129},
  url = {https://link.aps.org/doi/10.1103/PhysRevA.101.042129}
}

@article{Belfiglio:2025ofg,
    author = "Belfiglio, Alessio and Franzosi, Roberto and Luongo, Orlando",
    title = "{Geometric multipartite entanglement from gravitational particle production}",
    eprint = "2508.01658",
    archivePrefix = "arXiv",
    primaryClass = "gr-qc",
    month = "8",
    year = "2025"
}

@article{Vesperini:2023wks,
    author = "Vesperini, Arthur and Bel-Hadj-Aissa, Ghofrane and Capra, Lorenzo and Franzosi, Roberto",
    title = "{Unveiling the geometric meaning of quantum entanglement: Discrete and continuous variable systems}",
    eprint = "2307.16835",
    archivePrefix = "arXiv",
    primaryClass = "quant-ph",
    doi = "10.1007/s11467-024-1403-x",
    journal = "Front. Phys. (Beijing)",
    volume = "19",
    number = "5",
    pages = "51204",
    year = "2024"
}

@article{PhysRevD.31.1792,
  title = {Large-scale energy-density perturbations and inflation},
  author = {Lyth, D. H.},
  journal = {Phys. Rev. D},
  volume = {31},
  issue = {8},
  pages = {1792--1798},
  numpages = {0},
  year = {1985},
  month = {Apr},
  publisher = {American Physical Society},
  doi = {10.1103/PhysRevD.31.1792},
  url = {https://link.aps.org/doi/10.1103/PhysRevD.31.1792}
}

@article{Bunch:1978yq,
    author = "Bunch, T. S. and Davies, P. C. W.",
    title = "{Quantum Field Theory in de Sitter Space: Renormalization by Point Splitting}",
    doi = "10.1098/rspa.1978.0060",
    journal = "Proc. Roy. Soc. Lond. A",
    volume = "360",
    pages = "117--134",
    year = "1978"
}

@article{Danielsson:2003wb,
    author = "Danielsson, Ulf H. and Olsson, Martin E.",
    title = "{On thermalization in de Sitter space}",
    eprint = "hep-th/0309163",
    archivePrefix = "arXiv",
    reportNumber = "UUITP-17-03",
    doi = "10.1088/1126-6708/2004/03/036",
    journal = "JHEP",
    volume = "03",
    pages = "036",
    year = "2004"
}

@article{Greene:2005wk,
    author = "Greene, Brian and Parikh, Maulik and van der Schaar, Jan Pieter",
    title = "{Universal correction to the inflationary vacuum}",
    eprint = "hep-th/0512243",
    archivePrefix = "arXiv",
    reportNumber = "CU-TP-1136",
    doi = "10.1088/1126-6708/2006/04/057",
    journal = "JHEP",
    volume = "04",
    pages = "057",
    year = "2006"
}

@article{RevModPhys.96.045005,
  title = {Cosmological gravitational particle production and its implications for cosmological relics},
  author = {Kolb, Edward W. and Long, Andrew J.},
  journal = {Rev. Mod. Phys.},
  volume = {96},
  issue = {4},
  pages = {045005},
  numpages = {54},
  year = {2024},
  month = {Nov},
  publisher = {American Physical Society},
  doi = {10.1103/RevModPhys.96.045005},
  url = {https://link.aps.org/doi/10.1103/RevModPhys.96.045005}
}

@article{PhysRevD.105.123523,
  title = {Geometric corrections to cosmological entanglement},
  author = {Belfiglio, Alessio and Luongo, Orlando and Mancini, Stefano},
  journal = {Phys. Rev. D},
  volume = {105},
  issue = {12},
  pages = {123523},
  numpages = {9},
  year = {2022},
  month = {Jun},
  publisher = {American Physical Society},
  doi = {10.1103/PhysRevD.105.123523},
  url = {https://link.aps.org/doi/10.1103/PhysRevD.105.123523}
}

@article{PhysRevD.107.103512,
  title = {Inflationary entanglement},
  author = {Belfiglio, Alessio and Luongo, Orlando and Mancini, Stefano},
  journal = {Phys. Rev. D},
  volume = {107},
  issue = {10},
  pages = {103512},
  numpages = {16},
  year = {2023},
  month = {May},
  publisher = {American Physical Society},
  doi = {10.1103/PhysRevD.107.103512},
  url = {https://link.aps.org/doi/10.1103/PhysRevD.107.103512}
}

@article{PhysRevD.109.123520,
  title = {Superhorizon entanglement from inflationary particle production},
  author = {Belfiglio, Alessio and Luongo, Orlando and Mancini, Stefano},
  journal = {Phys. Rev. D},
  volume = {109},
  issue = {12},
  pages = {123520},
  numpages = {13},
  year = {2024},
  month = {Jun},
  publisher = {American Physical Society},
  doi = {10.1103/PhysRevD.109.123520},
  url = {https://link.aps.org/doi/10.1103/PhysRevD.109.123520}
}

@article{PhysRevLett.21.562,
  title = {Particle Creation in Expanding Universes},
  author = {Parker, L.},
  journal = {Phys. Rev. Lett.},
  volume = {21},
  issue = {8},
  pages = {562--564},
  numpages = {0},
  year = {1968},
  month = {Aug},
  publisher = {American Physical Society},
  doi = {10.1103/PhysRevLett.21.562},
  url = {https://link.aps.org/doi/10.1103/PhysRevLett.21.562}
}

@article{PhysRevD.39.389,
  title = {Particle creation in inhomogeneous spacetimes},
  author = {Frieman, Joshua A.},
  journal = {Phys. Rev. D},
  volume = {39},
  issue = {2},
  pages = {389--398},
  numpages = {0},
  year = {1989},
  month = {Jan},
  publisher = {American Physical Society},
  doi = {10.1103/PhysRevD.39.389},
  url = {https://link.aps.org/doi/10.1103/PhysRevD.39.389}
}

@article{Ford:2021syk,
    author = "Ford, L. H.",
    title = "{Cosmological particle production: a review}",
    eprint = "2112.02444",
    archivePrefix = "arXiv",
    primaryClass = "gr-qc",
    doi = "10.1088/1361-6633/ac1b23",
    journal = "Rept. Prog. Phys.",
    volume = "84",
    number = "11",
    year = "2021"
}

@article{Belfiglio:2025cst,
    author = "Belfiglio, Alessio and Luongo, Orlando and Mancini, Stefano",
    title = "{Quantum entanglement in cosmology}",
    eprint = "2506.03841",
    archivePrefix = "arXiv",
    primaryClass = "gr-qc",
    doi = "10.1016/j.physrep.2025.09.001",
    journal = "Phys. Rept.",
    volume = "1146",
    pages = "1--47",
    year = "2025"
}

@article{Verlinde:2016toy,
    author = "Verlinde, Erik P.",
    title = "{Emergent Gravity and the Dark Universe}",
    eprint = "1611.02269",
    archivePrefix = "arXiv",
    primaryClass = "hep-th",
    doi = "10.21468/SciPostPhys.2.3.016",
    journal = "SciPost Phys.",
    volume = "2",
    number = "3",
    pages = "016",
    year = "2017"
}

@article{Faulkner:2017tkh,
    author = "Faulkner, Thomas and Haehl, Felix M. and Hijano, Eliot and Parrikar, Onkar and Rabideau, Charles and Van Raamsdonk, Mark",
    title = "{Nonlinear Gravity from Entanglement in Conformal Field Theories}",
    eprint = "1705.03026",
    archivePrefix = "arXiv",
    primaryClass = "hep-th",
    doi = "10.1007/JHEP08(2017)057",
    journal = "JHEP",
    volume = "08",
    pages = "057",
    year = "2017"
}

\end{document}